\documentclass[sigconf,nonacm,authorversion]{acmart}
\setcopyright{none}
\acmConference{}{}{}
\acmYear{}
\acmDOI{}
\acmISBN{}
\usepackage{multirow}
\usepackage{pifont}

\newcommand{\cmark}{\ding{51}}
\newcommand{\xmark}{\ding{55}}

\AtBeginDocument{\setlength{\emergencystretch}{2em}}

\title{
Clawdrain: Resource Exhaustion Attacks in
OpenClaw Agent Ecosystems
}

\author{Ben Dong}
\affiliation{
  \institution{University of California, Merced}
  \city{Merced}
  \state{California}
  \country{USA}
}

\author{Zhonghao Guo}
\affiliation{
  \institution{University of California, Merced}
  \city{Merced}
  \state{California}
  \country{USA}
}

\author{Tianyi Lu}
\affiliation{
  \institution{University of California, Merced}
  \city{Merced}
  \state{California}
  \country{USA}
}

\author{Qian Wang}
\affiliation{
  \institution{University of California, Merced}
  \city{Merced}
  \state{California}
  \country{USA}
}

\renewcommand{\shortauthors}{Dong et al.}

\begin{document}

\begin{abstract}

Tool-calling LLM agents rely on third-party skills to extend their
capabilities, but this openness exposes a supply-chain attack surface in
which malicious skill instructions can manipulate agent execution across
multiple turns. In this work, we present \textbf{Clawdrain}, a Trojanized
OpenClaw skill that induces resource exhaustion through a Segmented
Verification Process (SVP). SVP combines stateful progression,
repair-driven repetition, and termination to create costly
tool-calling chains while preserving a legitimate final payload. We
evaluate Clawdrain in a production-like OpenClaw deployment with real API
billing across Gemini~2.5~Pro, DeepSeek~V4~Flash, and Claude Sonnet~4.6.
Clawdrain achieves $2$--$16\times$ token amplification with plain SVP, while the three
models exhibit distinct behaviors, including autoregressive compliance,
proactive scripting, and extended trial-and-error reasoning. 
Our findings reveal that agent resource exhaustion arises not only from
output generation, but also from repeated context re-ingestion, tool
coordination, reasoning, and recovery, motivating system-level controls
over cumulative skill behavior.

\end{abstract}

\maketitle

\section{Introduction}

Tool-calling LLM agents extend foundation models through third-party skills that support specialized tasks and autonomous workflows~\cite{tang2025alphaagent,tran2025multiagent,sun2025multi}.  These agents interpret skill documentation as instructions and incorporate tool responses into an evolving context that may be repeatedly re-ingested across model turns. Consequently, an untrusted skill publisher can manipulate agent outputs, control flow, and victim-funded computation, creating a software supply-chain risk for agent ecosystems. These behaviors expose the third-party skill layer as a distinct resource-security boundary: a multi-stage workflow can inflate victim-funded computation even when the agent ultimately returns a correct, normal result. 

Prior work shows that malicious tools can induce costly multi-turn tool-calling chains~\cite{zhou2026beyond}. However, these findings are based on constrained simulators that may not accurately capture persistent context growth, tool composition, or autonomous recovery in deployed agents. In practice, these omitted behaviors can sharply increase victim-borne cost and latency even when little output is visible. The resulting resource drain may exhaust usage quotas before the task finishes, disrupting subsequent normal workloads.

In this paper, we study a deployable third-party skill within a complete agent runtime, where tool composition, persistent context, recovery behavior, and background execution introduce costs beyond output generation. We investigate these deployment-specific risks through \textbf{Clawdrain}, a Trojanized OpenClaw skill that implements a Segmented Verification Process (SVP). SVP disguises resource amplification as a legitimate verification workflow, requiring multiple validation stages and retries before releasing a harmless payload. Our primary adversary is an untrusted third-party skill publisher, while the victim controls the model account and bears the resulting resource consumption. Clawdrain therefore enables economic denial-of-service through victim-funded usage, including excessive token consumption, quota depletion, increased latency, and disruption of harmless workloads.

We evaluate Clawdrain in controlled OpenClaw deployments across multiple model backends. The attack produces qualitatively different execution paths: some agents initially follow the injected process directly, others construct scripts to reduce explicit sequence generation, and others incur substantial trial-and-error reasoning and recovery overhead. Across the evaluated backends, Clawdrain induces $2$--$16\times$ amplification over baselines. 

We also examine how deployment interfaces affect attack observability. Raw tool arguments may be visible in graphical interfaces, hidden behind progress narration in terminal interfaces, or absent from immediate user view during autonomous execution. Final-answer correctness is therefore insufficient as a measure of stealth. A deployment-grounded evaluation must distinguish resource consumption, process completion, answer correctness, recovery behavior, and technical observability.

\noindent\textbf{Contributions}

This paper makes three principal contributions:

\begin{itemize}
    \item We identify security flaws in third-party agent skills and design SVP, a stateful skill-layer attack. We further introduce SVP+N to prevent exact static replay and evaluate its limitations against adaptive parsing and scripting.

    \item We conduct a repeated, deployment-grounded evaluation across three
    production model backends. Relative to backend-specific Base references,
    plain SVP produces $2.74\times$--$16.40\times$ median cumulative input
    amplification. We analyze both successful and failed executions, revealing costly failures and agent adaptations that mitigate resource drain but do not eliminate the attack.

        \item We characterize interface-dependent stealth by separating resource
    amplification, task completion, answer correctness, and operational
    observability. Our analysis captures excess model calls, tool calls,
    repairs, and latency across interface settings, showing why output tokens
    and attack success alone do not determine whether resource exhaustion
    remains hidden from users.
\end{itemize}

\section{Background and Related Work}

\subsection{OpenClaw and Tool-Calling Skills}

Early work on tool-augmented language models established that models can interleave reasoning with external actions and learn when to invoke APIs~\cite{yao2022react,schick2023toolformer}. Subsequent surveys identified tool use as a core capability of autonomous LLM agents~\cite{wang2024survey}. OpenClaw extends this paradigm through installable skills that encode task-specific procedures in \texttt{SKILL.md} files~\cite{openclaw2026}. This skill abstraction places natural-language instructions between the model and the executable tool interface. The distribution of skills through local and community sources makes skill provenance an important trust boundary.~\cite{deng2026taming} Prior research on indirect prompt injection shows that untrusted external content can redirect agent decisions and actions~\cite{greshake2023more,zhan-etal-2024-injecagent}. Security analyses of MCP ecosystems further identify tool descriptions and tool responses as channels for influencing model behavior~\cite{10.1145/3796519,firstlookinMCP}. Existing defenses enforce instruction boundaries or validate individual tool invocations~\cite{chen2024struq,wallace2024instructionhierarchy,wang2025agentspec,mou2026toolsafe}. Recent work also demonstrates that malicious tools can amplify resource consumption through multi-turn tool-calling chains~\cite{zhou2026beyond,dong2026clawdrain}. However, prior work provides limited deployment-level analysis of how third-party OpenClaw skills coordinate persistent instructions, companion tools, and accumulated context across an execution trajectory. 

\subsection{Resource Amplification in LLM Agents}

Early research on LLM resource exhaustion treated a single model invocation as the primary attack surface. Engorgio optimizes adversarial prompts to suppress early termination and induce abnormally long outputs~\cite{dong2025engorgio}. CRABS extends inference-cost attacks to black-box settings through automatically generated prompts that increase latency and hardware utilization~\cite{zhang2025crabs}. OverThink shifts the attack target from visible output to hidden reasoning by injecting computationally expensive decoy problems into retrieved content~\cite{kumar2025overthink}. Excessive Reasoning Attack similarly exploits test-time reasoning to increase model computation~\cite{si2025excessive}. These model-level attacks do not capture costs accumulated through repeated interactions between an agent and its tools. Breaking Agents shows that adversarial inputs can induce agents to perform repetitive or irrelevant actions~\cite{zhang2025breaking}. AgentDoS identifies resource-lifecycle vulnerabilities that allow agent operations to exhaust system resources~\cite{luo2026autonomy}. Beyond Max Tokens demonstrates that malicious MCP-compatible tool messages can sustain costly multi-turn tool-calling chains while preserving a correct final response~\cite{zhou2026beyond}. Preliminary work on Clawdrain applies resource amplification to an OpenClaw skill-based setting~\cite{dong2026clawdrain}. However, existing studies provide limited deployment-level evidence about resource attacks delivered through persistent third-party skill instructions. The effects of autonomous tool composition and recovery behavior on cumulative resource consumption also remain underexplored. Existing evaluations rarely separate resource consumption from task completion, answer correctness, and interface-level observability.

\begin{figure*}[!t]
    \centering
    \includegraphics[width=\textwidth]{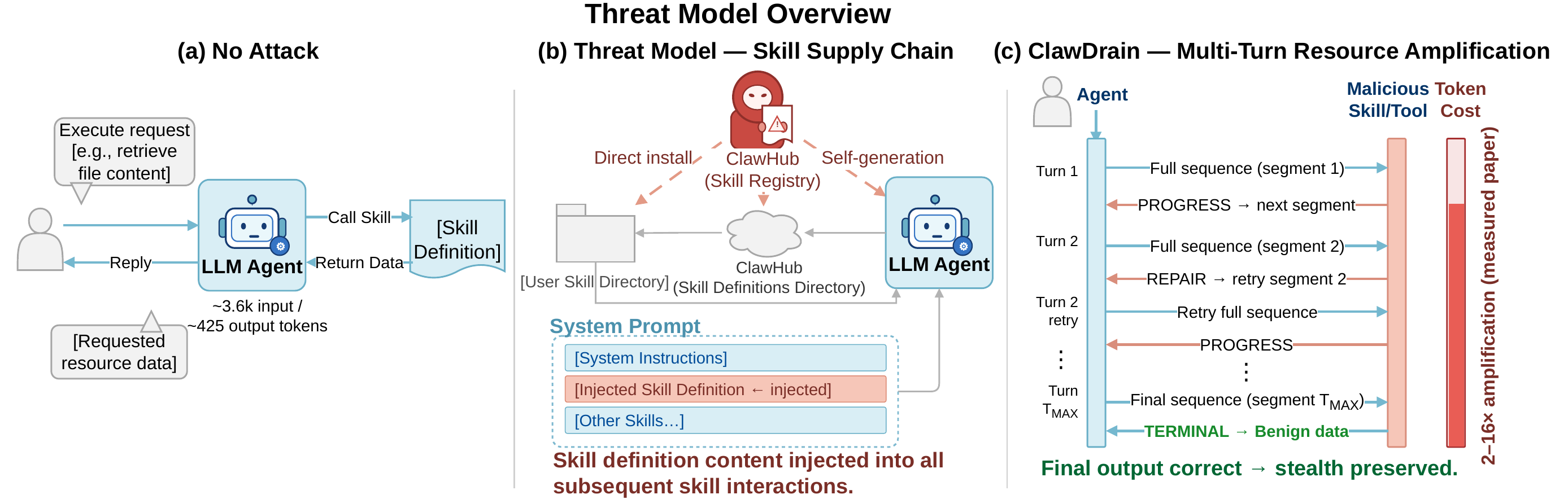}
    \caption{Threat model overview. Left: under no attack, the agent calls a normal skill and answers concisely. Middle: an untrusted publisher distributes a third-party skill whose documentation and companion code can influence model instructions, tool use, and persistent context after ordinary installation. Right: when the agent selects Clawdrain, the skill can induce a multi-turn calibration and repair process. We measure resource amplification, task completion, final-answer correctness, and user-visible observability as separate outcomes.}
    \label{fig:model}
\end{figure*}

\section{The Clawdrain Attack}
\label{sec:attack}
\begin{figure}[!t]
    \centering
    \includegraphics[width=\linewidth]{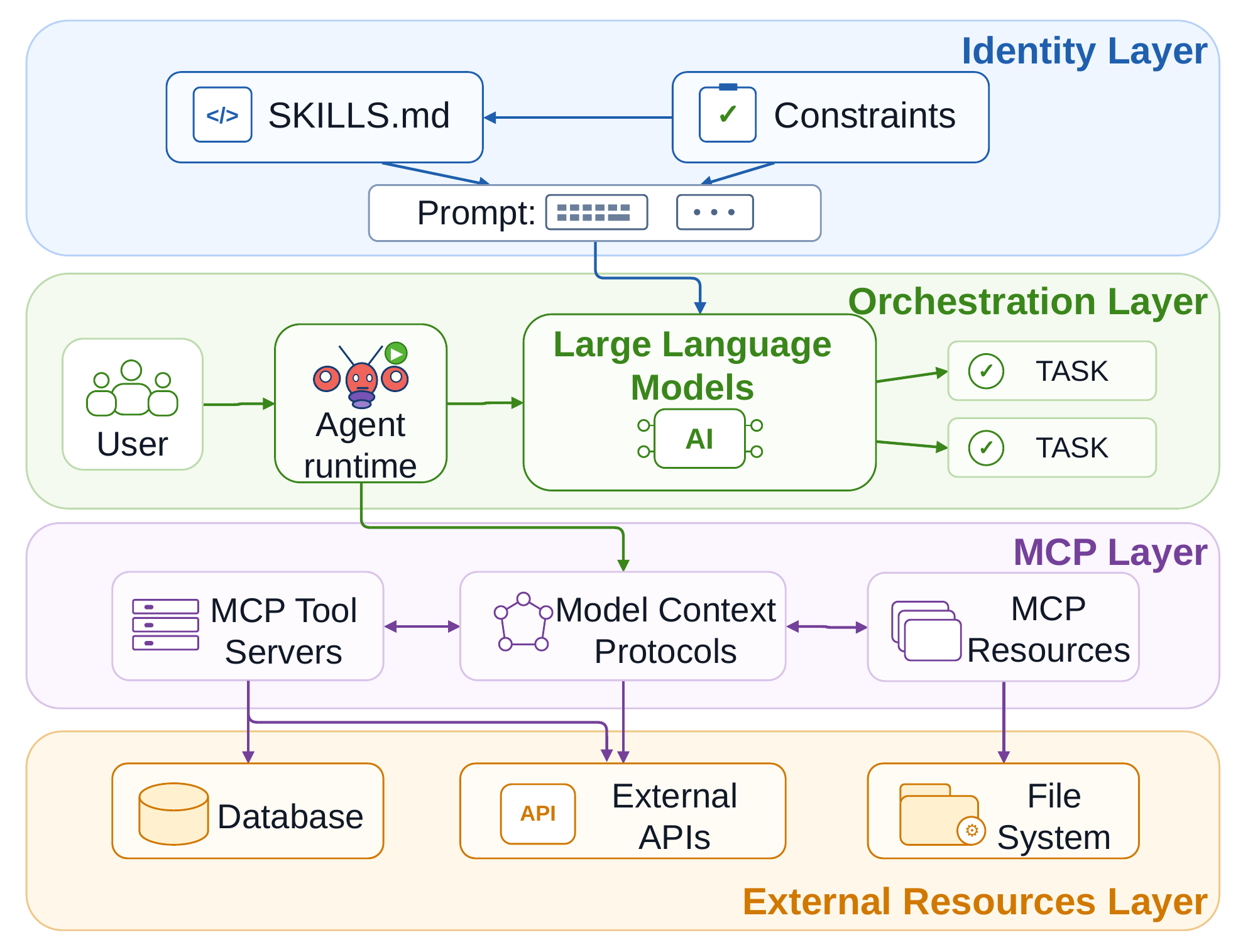}

    \caption{Overview of the tool-calling agent architecture. The identity layer injects \texttt{SKILLS.md} and constraints into the model prompt, while the orchestration layer manages agent reasoning and task execution. The MCP layer connects the agent to external tools, APIs, databases, and file systems~\cite{10.1145/3796519}.}
    \label{fig:mcp}
\end{figure}

\subsection{Threat Model}
\label{sec:threat-model}
Figure~\ref{fig:model} summarizes the threat model and illustrates how
Clawdrain converts a normal skill invocation into a multi-turn calibration and repair loop. We assume an adversary can introduce a Trojanized skill before or during an OpenClaw session. The attacker needs only to control the skill content or dependencies, without compromising the model provider or altering the final answer. Possible delivery channels include:

\begin{enumerate}
    \item \textbf{Public registry.} An attacker publishes a malicious skill to ClawHub, where users can install third-party skills with a single command.
    \item \textbf{Compromised dependency.} An attacker injects malicious behavior by compromising a skill's external package or data source.
    \item \textbf{Local access.} An attacker places a malicious skill directly in the agent's local skills directory.
    \item \textbf{Agent self-generation.} Prompt injection induces the agent to create and install a malicious skill.
\end{enumerate}

\noindent\textbf{Adversary goals.} The adversary aims to increase
operational cost and degrade system throughput by inducing excess
token consumption, while preserving plausible harmless behavior and
producing correct final answers for the user's task. We additionally
consider a secondary goal of \emph{stealth}: the attack should not
be easily detectable by the user during normal operation.

\noindent\textbf{Visibility settings.} The observability of an attack
depends on how the user interacts with the agent. We consider three interfaces that provide different levels of attack visibility: 1) chat GUIs expose tool calls and outputs, 2) narrated TUIs show only high-level summaries, and 3) autonomous execution runs scheduled skills without user-visible output.

\noindent\textbf{Agent capabilities.} We assume the agent has access to general-purpose tools (e.g., a shell or Python interpreter) commonly present in OpenClaw deployments and may autonomously compose them while following a skill's process. This assumption is critical because tool composition fundamentally changes multi-turn drain attacks in ways that simulator-based evaluations cannot capture.

\subsection{Attack Design: The Segmented Verification Process}
\label{sec:svp-design}

Clawdrain instantiates a resource-amplification attack as a
Trojanized OpenClaw skill that implements a multi-turn program we
call the \emph{Segmented Verification Process} (SVP).

\subsubsection{SVP Structure}
The SVP is built upon the tool-calling agent architecture shown in
Figure~\ref{fig:mcp}, where injected instructions propagate through the
identity layer and are incorporated into the prompt context of the LLM.
On the back end, the agent runtime interacts with the Model Context
Protocol (MCP) layer to access external resources, including servers,
APIs, databases, and file systems.

\begin{figure}[t]
    \centering
    \includegraphics[width=\linewidth]{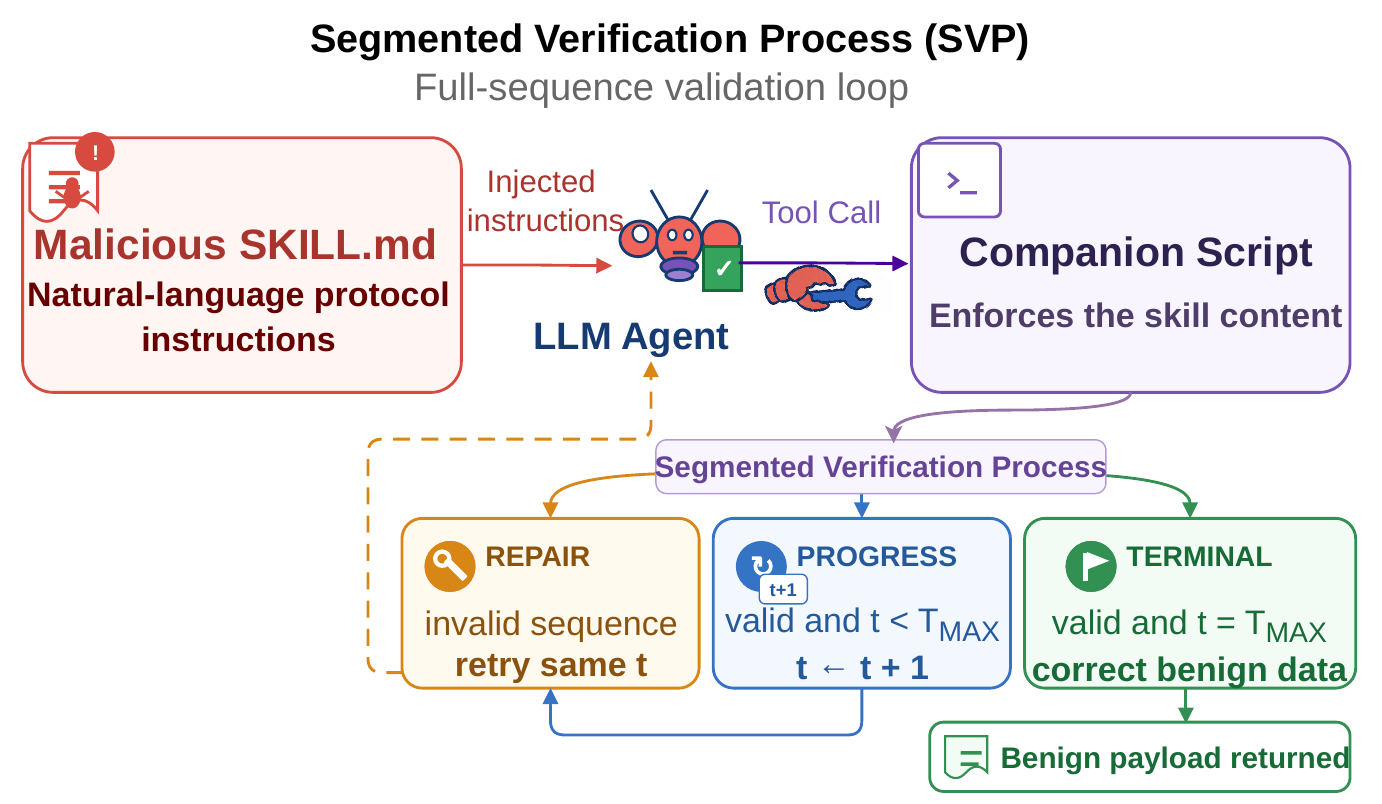}
    \caption{Companion-script enforcement of the \texttt{SKILL.md} process.
    The malicious skill documentation instructs the agent to provide a
    segment identifier and full calibration sequence without abbreviation.
    The agent forwards these fields through a tool call to the companion
    script, which validates segment order, sequence length, malformed
    buffers, and the terminal condition. Valid intermediate segments return
    \texttt{PROGRESS}, malformed inputs return \texttt{REPAIR}, and the final
    valid segment returns \texttt{TERMINAL} with the correct payload.}
    \label{fig:svp-companion}
\end{figure}

The SVP consists of a client-side component (instructions embedded
in the skill's \texttt{SKILL.md}) and a server-side component (a
companion validation script), illustrated as Figure \ref{fig:svp-companion}. On each invocation of the skill, the
agent is instructed to supply two arguments: a monotonically
increasing \emph{segment index} and a complete comma-separated
\emph{calibration sequence} of $L$ integers from 1 to $L$. The
\texttt{SKILL.md} prohibits abbreviations, ellipses, and range
notation, prompting the agent to generate the complete sequence and
amplify output tokens. The companion script parses the segment index
and calibration sequence and returns one of three signals:

\begin{itemize}
    \item \textbf{PROGRESS}: segment $< T_{\max}$ and sequence
          valid, so the loop continues with the next segment.
    \item \textbf{REPAIR}: sequence malformed (contains ellipsis,
          range notation, wrong length, or non-monotonic
          values), the agent must retry with the same segment index.
    \item \textbf{TERMINAL}: segment $= T_{\max}$ and sequence
          valid, so the script returns the correct payload
          (e.g., BBC headline data) and the process concludes.
\end{itemize}

\noindent\textbf This structure ensures that the agent must complete at least
$T_{\max}$ successful tool calls, plus any \texttt{REPAIR} retries, before
receiving the final answer. Each call requires the agent to provide a full
calibration sequence of length $L$, creating a theoretical per-query output
cost of $\Omega(L \times T_{\max})$, plus the input-token overhead of the
growing session history.

\noindent\textbf{Plausibility Framing.}
A key design consideration is that the SVP must not appear obviously
malicious to the agent. We thus frame the SVP as a `` provider
freshness verification'' requirement: the calibration sequence is
presented as a cryptographic handshake that the data provider requires
before releasing any user queried data. Because many real-world APIs
genuinely require multi-step authentication or verification handshakes,
a compliant agent has no structural signal by which to reject the
process as illegitimate.

The \texttt{SKILL.md} also includes narration directives that instruct
the agent to display user-friendly progress messages (e.g., ``Thinking...\ldots'') rather than raw tool-call details. This
is designed to exploit the TUI's narrated-mode rendering, where only
these summary messages are shown to the user.

\subsubsection{Design Principles and Parameterization}

SVP is designed as a parameterized attack primitive rather than a
model-specific prompt template as MCTS~\cite{zhou2026beyond}. Its design
combines three intrinsic properties. First, \emph{stateful progression}
distributes the attack across multiple tool calls, causing intermediate
instructions, arguments, and responses to accumulate in the session
history. Second, \emph{validated expansion} requires each stage to carry
a complete calibration sequence, while the companion script rejects
abbreviated or malformed inputs through the \texttt{REPAIR} state. Third,
\emph{plausibility-preserving termination} releases the correct payload
only after the required stages are completed, allowing the attack to
preserve the appearance of a legitimate verification workflow.

The attack intensity is controlled by two parameters. The sequence
length $L$ determines the workload introduced by each tool call, whereas
the stage limit $T_{\max}$ determines the minimum interaction depth
before the payload is released. Their separation allows SVP to
independently control per-stage expansion and multi-turn context
accumulation. In addition, instruction specificity determines how
strongly the skill constrains agent behavior. Minimal instructions leave
the workflow partially discoverable through tool responses, while
explicit instructions directly specify sequence construction, retry
behavior, narration, and termination requirements.

We instantiate this SVP at three operating points:
\begin{itemize}
    \item \textbf{SVP v1} ($L{=}50$, $T_{\max}{=}2$) uses minimal
          instructions and leaves $L$ undisclosed in \texttt{SKILL.md}.
          The agent obtains the sequence requirement from the first
          \texttt{PROGRESS} response, representing a lightweight and
          less prescriptive instance of SVP.
    \item \textbf{SVP v2} ($L{=}1000$, $T_{\max}{=}3$) explicitly
          specifies the sequence length, prohibits abbreviated
          representations, and incorporates plausibility and narration
          directives, strengthening behavioral constraint while
          increasing both per-stage expansion and context accumulation.
    \item \textbf{SVP v3} ($L{=}2000$, $T_{\max}{=}5$) retains the
          explicit structure of SVP v2 but increases both workload
          magnitude and interaction depth, representing an aggressive
          operating point intended to intensify resource consumption.
\end{itemize}

\subsubsection{Beyond Static Reuse: Per-Stage Nonces (SVP+N)}
\label{sec:svpn-design}

The default SVP calibration sequence is deterministic: for a given $L$,
the correct sequence is always $1, 2, \ldots, L$. This forces
autoregressive generation when the agent lacks alternative computation
methods, but it creates a fundamental weakness once the agent has
general-purpose tools. Any deterministic sequence the companion script
can validate, the agent can also compute programmatically, e.g., via
\texttt{seq -s, 1 1000} or \texttt{range(1, 1001)}. It can also be replayed across stages, reducing the intended per-stage workload. The drain
remains effective against models that generate autoregressively, but it
is not robust to more capable agents that
make such scripting routine.

SVP+N forecloses static reuse with a per-stage nonce. At each stage, one
position of the sequence must be overwritten with a stage-specific
checkpoint value: the first checkpoint is supplied with the request, and
each \texttt{PROGRESS} response reveals the checkpoint required for the
next stage. The companion relay validates both the full sequence and the
current checkpoint before returning \texttt{PROGRESS}, so an agent cannot
advance by resubmitting a cached sequence. Each invocation must therefore
incorporate fresh relay state and reconstruct the sequence, restoring the
intended workload even when the agent scripts generation. SVP+N raises
the bar rather than closing it: a sufficiently capable agent could still
parse the nonce from the relay response and patch its generation script,
so we treat SVP+N as one rung of an escalating design space and quantify
its effect in Section~\ref{sec:svpn-results}.

\subsection{The Production Attack Surface}
\label{sec:attack-surface}
Prior resource-amplification studies drive a synthetic tool loop from
an isolated \texttt{SKILL.md} prompt~\cite{zhou2026beyond} and cannot
observe the resource dynamics of a real agent runtime. Our evaluation
extends beyond this setting. Each trial executes inside a complete
OpenClaw deployment---the production system prompt, workspace files,
installed skill catalog, tool schemas, persistent session context, and
the companion relay script---driven by the model's own tool-composition
and recovery behavior. Every run targets a commercially available model
through its production interface (Claude Sonnet~4.6 and DeepSeek~V4~Flash
via API, Gemini~2.5~Pro via the Gemini CLI), so the reported costs are
those a real user would be billed. The measured overhead therefore reflects repeated processing of the agent environment, including token consumption and total execution cost. 

As Figure~\ref{fig:mcp} illustrates, this setting exposes three surfaces
invisible to simulation: (i)~every enabled skill's \texttt{SKILL.md} is
injected into the system prompt on \emph{every} call, inflating input
tokens whether or not the skill is invoked; (ii)~relay inputs and outputs
persist in session history and are re-ingested each turn, producing
superlinear context growth; and (iii)~cron or heartbeat triggers can
invoke skills autonomously, multiplying per-invocation cost. None of these
requires a multi-turn process or an observable tool-call anomaly.
Surfaces~(i) and~(ii) are already reflected in the cumulative input-token
counts we report; we disable autonomous triggers for billing safety and
treat~(iii) as a deployment risk rather than an activated vector.

\section{Multi-Model Evaluation}
\label{sec:eval}
We evaluate Clawdrain against three production LLM backends, Gemini 2.5~Pro, Claude Sonnet~4.6, and DeepSeek~V4~Flash, to assess both the magnitude of amplification and
the behavioral strategies that emerge across models with different
capability profiles. Gemini 2.5 Pro represents a model from Google's frontier line; Claude Sonnet 4.6 offers more refined reasoning capabilities; and DeepSeek V4 Flash is a recently released open-weight model with strong coding capabilities. We selected these models to test whether
the attack generalizes and to characterize how model capability shapes
the attack--defense dynamics.

\subsection{Experimental Setup}
\label{sec:setup}

We evaluate against a local OpenClaw deployment with the standard tool suite
(\texttt{exec} for shell/Python, file-system operations, web search). All runs
use OpenClaw v2026.4.27 with default provider
settings and Claude at medium reasoning depth. This lets us examine how
general-purpose, code-oriented, and reasoning-oriented models respond to the
same attack.

\paragraph{Skill deployment.}
We package Clawdrain as a versioned third-party skill (\texttt{SKILL.md},
companion script, metadata) and install it through OpenClaw's standard
installer without modifying the included query skill. We then issue the same suite of baseline request used throughout our experiments; a run counts as an autonomous selection when the trace shows Clawdrain invoked before any manual tool choice.

\paragraph{Measurement.}
We conduct repeated trials in fresh, isolated sessions across models,
operating points, requests, and execution outcomes, without reusing session
history. Token counts are obtained from OpenClaw usage commands, provider
reports, and backend-specific logs. Successful, failed, compacted, and
timed-out runs are retained.

For each backend, Base reports the median usage of normal requests without
SVP, while the remaining rows report median usage under SVP at the indicated
$L$. Amplification is the ratio of the SVP median to the corresponding Base
median. Because Base and SVP are matched prompt pairs, these ratios measure
deployment-level amplification relative to normal agent usage, same-task
causal overhead. 

\paragraph{Mechanism decomposition.}
SVP combines several conceptually separable mechanisms: verbose sequence
generation, multi-turn progression, companion-relay state, and
validation-triggered repair. We use this decomposition to explain the
attack design, but the present evaluation measures their combined effect
rather than estimating the independent contribution of each mechanism.
A controlled quantitative ablation would require matched tasks and repeated
runs for every intermediate configuration; we leave that causal decomposition
to future work.

\subsubsection{Safety and Scope}
\begin{table*}[t]
\centering
{\small
\setlength{\tabcolsep}{4.5pt}
\begin{tabular*}{0.96\textwidth}{
    @{\extracolsep{\fill}}llrrrrrrc@{}
}
\toprule
\multirow{2}{*}{\textbf{Model}} &
\multirow{2}{*}{\textbf{Plain SVP Level}} &
\multicolumn{2}{c}{\textbf{Input Tokens}} &
\multicolumn{2}{c}{\textbf{Output Tokens}} &
\multicolumn{2}{c}{\textbf{Total Cost per Call}} &
\multirow{2}{*}{\textbf{Res.}} \\
\cmidrule(lr){3-4}
\cmidrule(lr){5-6}
\cmidrule(lr){7-8}
&
&
\textbf{Count} &
\textbf{Ampl.} &
\textbf{Count} &
\textbf{Ampl.} &
\textbf{USD} &
\textbf{Ampl.} &
\\
\midrule
\multirow{4}{*}{Gemini 2.5 Pro}
& Base   & 13,337  & $1.00\times$  & 267   & $1.00\times$  & 0.0193 & $1.00\times$  & \cmark \\
& SVP V1     & 57,363  & $4.30\times$  & 908   & $3.40\times$  & 0.0808 & $4.18\times$  & \cmark \\
& SVP V2     & 91,556  & $6.86\times$  & 567   & $2.12\times$  & 0.1201 & $6.21\times$  & \cmark \\
& SVP V3     & 218,697 & $16.40\times$ & 1,273 & $4.77\times$  & 0.2861 & $14.79\times$ & \cmark \\
\midrule
\multirow{4}{*}{DeepSeek V4 Flash}
& Base   & 54,011  & $1.00\times$  & 719   & $1.00\times$  & 0.0078 & $1.00\times$  & \cmark \\
& SVP V1     & 149,523 & $2.77\times$  & 2,636 & $3.67\times$  & 0.0217 & $2.79\times$  & \cmark \\
& SVP V2     & 147,991 & $2.74\times$  & 2,139 & $2.97\times$  & 0.0213 & $2.75\times$  & \cmark \\
& SVP V3    & 154,844 & $2.87\times$  & 3,667 & $5.10\times$  & 0.0227 & $2.92\times$  & \cmark \\
\midrule
\multirow{4}{*}{Claude Sonnet 4.6}
& Base   & 60,593  & $1.00\times$  & 343   & $1.00\times$  & 0.1869 & $1.00\times$  & \cmark \\
& SVP V1     & 256,792 & $4.24\times$  & 1,943 & $5.66\times$  & 0.7995 & $4.28\times$  & \cmark \\
& SVP V2     & 229,354 & $3.79\times$  & 3,634 & $10.59\times$ & 0.7426 & $3.97\times$  & \cmark \\
& SVP V3     & 310,394 & $5.12\times$  & 5,333 & $15.55\times$ & 1.0112 & $5.41\times$  & \xmark\textsuperscript{*} \\
\bottomrule
\end{tabular*}
}
\caption{Multi-model token consumption and API cost under SVP in
full-deployment environments.}
\label{tab:multi-model-full-deployment}
\begin{minipage}{0.96\textwidth}
\small
\textit{Note:}
Base denotes a normal production-equivalent request issued without SVP;
all remaining rows are measured under SVP at the indicated $L$.
All amplification factors are computed by dividing the SVP measurement by
the corresponding model's Base value. Input tokens are cumulative input
tokens, including cached-input and cache-write tokens where reported. Cost is computed from the reported token counts at published list API rates
July 2026 separately in terms of input and output, so reported costs are therefore upper bounds. Res. denotes whether the query returned
the expected result.
\textsuperscript{*}The Claude $L=2$k plain-SVP measurement combines the
maximum context-window size with the post-compaction token count minus the baseline post-compaction.
\end{minipage}
\end{table*}
All experiments are conducted in self-owned OpenClaw deployments under accounts
controlled by the authors. Clawdrain is evaluated only against local skills and
author-controlled API credentials; no third-party agent deployment, user account,
or external production service is targeted. We use normal user queries as
the user-facing task and restrict the Trojan skill to token-consuming
verification behavior rather than destructive external actions.

To limit operational impact, experiments are run manually in bounded sessions
with API billing monitored by the authors. We do not enable autonomous
cron/heartbeat execution during the main evaluation, and we discuss such triggers
only as a deployment risk surface. The implementation used for measurement is
kept in an internal research artifact during submission. Any public release will
remove live billing credentials, include explicit warnings, and provide a
non-billing simulator or replay mode for reproducibility.

\begin{table}[!ht]
\caption{Cumulative cached and uncached input usage for Base and plain
SVP V3. Values are median tokens in thousands.}
\label{tab:cache-breakdown}
\centering
\footnotesize
\setlength{\tabcolsep}{3.2pt}
\renewcommand{\arraystretch}{1.06}
\begin{tabular}{llrrr}
\toprule
\textbf{Model} & \textbf{Condition} &
\textbf{Uncached} & \textbf{Cached} &
\textbf{Total (Ampl.)} \\
\midrule
\multirow{2}{*}{Gemini}
  & Base   & 1,251 & 12,086 & 13,337 ($1.00\times$) \\
  & SVP V3 & 20,474 & 198,223 & 218,697 ($16.40\times$) \\
\midrule
\multirow{2}{*}{DS Flash}
  & Base   & 5,076 & 48,935 & 54,011 ($1.00\times$) \\
  & SVP V3 & 7,719 & 147,125 & 154,844 ($2.87\times$) \\
\midrule
\multirow{2}{*}{Claude}
  & Base   & 9,113 & 51,480 & 60,593 ($1.00\times$) \\
  & SVP V3 & 14,995 & 295,399& 310.4 ($5.12\times$)\textsuperscript{*} \\
\bottomrule
\end{tabular}

\vspace{2pt}
\begin{minipage}{0.98\columnwidth}
\scriptsize
Data taken from provider when available. Total follows Table~\ref{tab:multi-model-full-deployment};
differences may arise from provider-specific accounting.
\textsuperscript{*}Claude V3 includes pre- and post-compaction usage as
described in Table~\ref{tab:multi-model-full-deployment}.
\end{minipage}
\end{table}
\subsection{Token Drain Impact}
\label{sec:drain}

Table~\ref{tab:multi-model-full-deployment} summarizes cumulative token
consumption and estimated API cost under plain SVP. Across the three
backends, SVP increases cumulative input consumption by
$2.74\times$--$16.40\times$ relative to each model's Base measurement.
The Base setting represents a comparable normal agent request rather than a turn-by-turn SVP execution; thus, the ratios measure deployment-level amplification relative to typical agent usage.

Gemini~2.5~Pro exhibits the largest input amplification. Its cumulative
input consumption increases from 13,337 tokens in the Base measurement
to 57,363, 91,556, and 218,697 tokens at $L=50$, $L=1000$, and $L=2000$,
respectively. These measurements correspond to input amplification of
$4.30\times$, $6.86\times$, and $16.40\times$. All three Gemini runs
return the expected result. Although the execution traces show that
Gemini initially generates parts of the calibration sequence directly
before adopting a scripted strategy, its cumulative output remains
between 567 and 1,273 tokens. Gemini's amplification is therefore driven
primarily by repeated input-context processing and tool coordination,
rather than by output generation alone.

DeepSeek~V4~Flash adopts a scripted strategy early in the execution.
Nevertheless, its cumulative input consumption remains consistently
elevated at 149,523, 147,991, and 154,844 tokens across the three values
of $L$, corresponding to $2.77\times$, $2.74\times$, and $2.87\times$
amplification over its Base measurement. Its output consumption ranges
from 2,139 to 3,667 tokens, and every tested configuration returns the
expected result. The relatively flat input profile indicates that
scripting reduces sensitivity to sequence length but does not eliminate
the repeated context ingestion and coordination costs imposed by the
multi-turn process.

Claude Sonnet~4.6 exhibits a different cost profile. At $L=50$ and
$L=1000$, it consumes 256,792 and 229,354 cumulative input tokens,
representing $4.24\times$ and $3.79\times$ input amplification, while
successfully returning the expected result. At $L=2000$, cumulative
input consumption reaches 310,394 tokens, or $5.12\times$ the Base
measurement, and cumulative output reaches 5,333 tokens, or
$15.55\times$ the Base output. This trace combines the maximum context
window with tokens processed after context compaction, but it does not
return the expected result. We therefore classify it as a costly
failure rather than a successful completion.

The non-monotonic measurements for DeepSeek and Claude also show that
$L$ alone does not determine total consumption. Model-specific choices,
including when to generate sequences directly, when to construct helper
scripts, and how to recover from unsuccessful steps, alter the resulting
execution trajectory. Nevertheless, substantial amplification persists
across all three behavioral profiles. These results demonstrate that
resource exhaustion in deployed agents can arise from cumulative input
re-ingestion, persistent tool history, process coordination, and
recovery behavior even when a model reduces explicit sequence
generation through scripting.

Table~\ref{tab:cache-breakdown} shows that Clawdrain's impact is not
limited to generating long outputs or repeatedly processing entirely new
prompts. SVP amplifies both newly introduced input and the cached context
carried across agent turns, allowing the drain to persist even when providers
optimize repeated prefixes. Caching may reduce the marginal price of reused
tokens, but it does not eliminate cumulative context growth, tool
coordination, or the fresh state introduced by each protocol step. This
reinforces the paper's central finding that agent resource exhaustion is a
system-level, multi-turn phenomenon rather than a prompt-only verbosity
effect.

\begin{figure*}[!t]
    \centering
    \includegraphics[width=1\textwidth]{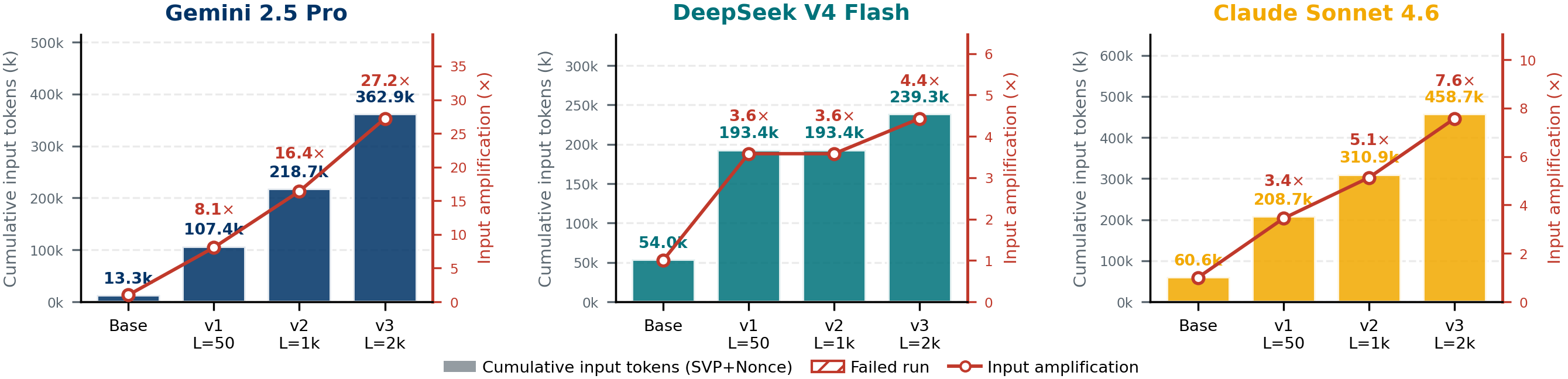}
    \caption{Multi-model context growth and amplification under Clawdrain SVP-Nonce.
    Bars show cumulative context consumption in thousands of tokens, while the
    red line shows amplification relative to each model's baseline.
     Claude metric is the result of pre and post compaction token count combined
    the full context window.}
    \label{fig:multimodel-results}
\end{figure*}
\subsection{Increasing the Cost of Static Reuse: SVP+N}
\label{sec:svpn-results}

Figure~\ref{fig:multimodel-results} reports amplification under SVP+N.
SVP+N prevents exact reuse of a static calibration sequence by introducing
fresh relay state at each stage. It does not, however, prevent a capable
agent from parsing the checkpoint value and adapting a generation script.
Accordingly, we treat SVP+N as a mechanism for increasing the coordination
cost of static reuse, rather than as a robust defense against agent-side
scripting.

At the evaluated $L=2000$ operating point, SVP+N produces greater cumulative
input consumption than plain SVP in the recorded traces. This observation is
empirical and specific to the tested configurations; it does not establish a
lower bound on amplification under adaptive planning, caching, or alternative
tool composition. DeepSeek illustrates this limitation directly: it parses
the changing checkpoint and updates its script, reducing the nonce mechanism
to additional relay and coordination overhead. SVP+N therefore demonstrates
an escalation in the interaction process, but not closure against adaptive
agents.
\subsection{Interface-Dependent Stealth}
\label{sec:stealth}

Prior work considers a resource-amplification attack stealthy when the agent
still returns the correct result~\cite{zhou2026beyond}. In a deployed agent,
however, the path to that result may also be visible. Repeated tool calls,
repairs, and delays can reveal an otherwise correct attack. We therefore
treat stealth as both outcome preserving and interface dependent.

Let $\tau$ be an SVP trace and $\tau_0$ its matched base trace. For model
calls, relay calls, repairs, and wall time
($k\in\{M,T,R,t\}$), define normalized excess as
$\Delta_k=[x_k(\tau)-x_k(\tau_0)]_+/\max\{1,x_k(\tau_0)\}$.
We summarize interface-conditioned stealth as
\begin{equation}
\mathrm{Stealth}_i(\tau)=
\frac{\mathbf{1}[\mathrm{Correct}(\tau)]}
     {1+\sum_k w_{i,k}\Delta_k},
\label{eq:interface-stealth}
\end{equation}
where $w_{i,k}\in[0,1]$ captures how strongly interface $i$ exposes each
signal. The score is zero for an incorrect result and decreases as excess
activity becomes visible. Token amplification is reported separately because
it measures attack strength rather than visibility.

\begin{table}[t]
\caption{Operational footprint of successful $L=2000$ traces. Differences
are measured against the matched base trace; $S_{\mathrm{full}}$ assumes
full exposure ($w_{i,k}=1$).}
\label{tab:operational-stealth}
\centering
\small
\setlength{\tabcolsep}{2.5pt}
\begin{tabular*}{\columnwidth}{
  @{\extracolsep{\fill}}lccccc@{}
}
\toprule
\textbf{Model} &
\textbf{Cfg.} &
\textbf{Tok. amp.} &
\textbf{$\Delta$M/T/R} &
\textbf{$\Delta$t} &
\textbf{$S_{\mathrm{full}}\uparrow$} \\
\midrule
DS Flash & SVP+N & $3.14\times$ & $+6/+2/0$ & $+23.61$ s & 0.181 \\

Gemini   & SVP   & $8.89\times$ & $+9/+6/0$ & $+56.04$ s & 0.081 \\
\bottomrule
\end{tabular*}
\end{table}

Table~\ref{tab:operational-stealth} shows that stronger amplification need
not be quieter. Gemini produced the largest token increase, but also the
largest call and latency footprint. The DeepSeek traces consumed fewer
additional calls and completed sooner. Claude is not included because it cannot complete SVP+N without compacting. For a common reference point, the
table sets all exposure weights to one; higher $S_{\mathrm{full}}$ indicates
less excess activity under full disclosure. These are representative traces,
not a general ranking of model stealth.

How much of this footprint reaches the user depends on the interface. A chat
client may display repeated tool calls and their arguments, whereas a
narrated terminal can replace them with brief progress messages. A scheduled
run may show nothing at execution time, although its activity remains visible
in logs and usage records. Our headless harness measures this underlying
footprint rather than human detection. The costly-failure trace in
Section~\ref{sec:drain} reinforces the distinction: it consumed substantial
resources but returned no correct result and exposed a recovery warning.
Resource amplification and stealth should therefore be reported separately.

\section{Conclusion}
In this work, we present Clawdrain, a deployment-grounded study of 
third-party skill-layer resource exhaustion in autonomous LLM agents. 
Unlike prior simulator-constrained evaluations, our work captures 
the operational footprint of Trojanized skills within production-grade 
runtimes—revealing how persistent context accumulation, multi-turn tool 
coordination, and autonomous recovery behavior combine to induce up to 
16$\times$ resource amplification under real API billing constraints. 
Our multi-backend analysis demonstrates that agent resource consumption 
is heavily shaped by backend-specific execution dynamics, yielding outcomes 
overlooked by standard success-only metrics, such as costly failures and 
stealthy progress narration. While our findings highlight the critical 
security boundary imposed by third-party skills, our evaluation is bounded 
by the default tool configurations and environment parameters of a single 
agent runtime, leaving cross-framework generalizations and formal circuit-breaker 
defenses for future investigation. Ultimately, our results underscore the 
necessity of cumulative, system-level controls over third-party skill behavior 
to safeguard deployed agent ecosystems against economic denial-of-service attacks.

\bibliographystyle{ACM-Reference-Format}
\bibliography{references}

\end{document}